\documentclass[12pt,a4paper]{article}

\usepackage{fullpage}
\usepackage{authblk}
\usepackage{amsthm}
\usepackage{amsmath}
\usepackage{amsfonts}
\usepackage{microtype}
\usepackage{hyperref}
\usepackage{graphicx}
\usepackage{xcolor}

\newcommand{\ZZ}{\mathbb{Z}}

\DeclareMathOperator{\weight}{weight}
\DeclareMathOperator{\dist}{query}
\renewcommand{\O}{\mathcal{O}}

\title{Shadoks Approach to\\ Low-Makespan Coordinated Motion Planning}

	\author[1]{Loïc Crombez}
	\author[2]{Guilherme D. da Fonseca}
	\author[1]{Yan Gerard}
    \author[2]{Aldo Gonzalez-Lorenzo}
    \author[1]{Pascal Lafourcade}
    \author[1]{Luc Libralesso}

    \affil[1]{LIMOS, Université Clermont Auvergne}
    \affil[2]{LIS, Aix-Marseille Université}

\begin{document}

\maketitle

\begin{abstract}
This paper describes the heuristics used by the \texttt{Shadoks}\footnote{The team name comes from the animated television series Les Shadoks \url{https://en.wikipedia.org/wiki/Les_Shadoks}.} team for the CG:SHOP 2021 challenge.  This year's problem is to coordinate the motion of multiple robots in order to reach their targets without collisions and minimizing the makespan. It is a classical multi agent path finding problem with the specificity that the instances are highly dense in an unbounded grid. Using the heuristics outlined in this paper, our team won first place with the best solution to 202 out of 203 instances and optimal solutions to at least 105 of them. The main ingredients include several different strategies to compute initial solutions coupled with a heuristic called Conflict Optimizer to reduce the makespan of existing solutions.
\end{abstract}

\section{Introduction} \label{s:intro}
We explain some heuristics used by the \texttt{Shadoks} team to win first place in the CG:SHOP 2021 challenge that considers a coordinated motion planning problem in the two-dimensional grid $\ZZ^2$. The goal is to move a set of $n$ labeled unit squares called \emph{robots} between given start and target grid cells without collisions.

More formally, the input consists of a set of \emph{obstacles} $O$ and a set of $n$ \emph{robots} $R=\{r_1,\ldots,r_n\}$. Each obstacle is a lattice point and each robot $r_i$ is a pair of lattice points $s_i,d_i$ respectively called \emph{start} and \emph{target}. A path $P_i$ of length $m$ is a sequence of $m+1$ lattice points $p_i(t)$ for a time $t = 0, \ldots, m$. A solution of makespan $m$ is an assignment of \emph{paths} $P_i$ of length $m$ to each robot $r_i$ satisfying the following constraints.
\begin{enumerate}
\item $p_i(0) = s_i$ and $p_i(m) = d_i$ for $1 \leq i \leq n$,
\item $\|p_i(t) - p_i(t-1)\| \leq 1$ for $1 \leq i \leq n$ and $1 \leq t \leq m$,
\item $p_i(t) \notin O$ for $1 \leq i \leq n$ and $0 \leq t \leq m$,
\item (\emph{collision constraint}) $p_i(t) \neq p_j(t)$ for $i \neq j$ and $0 \leq t \leq m$, and 
\item (\emph{overlap constraint}) if $p_i(t) = p_j(t-1)$, then $p_i(t) - p_i(t-1) = p_j(t) - p_j(t-1)$. This constraint comes from the robots being square shaped, in order to avoid corner collisions in the continuous movement of the robots.
\end{enumerate}

The objective of the problem is to minimize the makespan\footnote{The challenge also considered the objective of minimizing the sum of the distances, but we did not optimize our solutions for this version of the problem.} $m$. A trivial lower bound is obtained by ignoring constraints (4) and (5) and finding shortest paths.
For more details about the problem, see the overview~\cite{overview} and the related paper~\cite{coordinated2019}.

The challenge CG:SHOP 2021 provided $203$ instances containing between $10$ and $9000$ robots, out of which $202$ of our solutions were the best ones among all the $17$ teams who participated. To our surprise, we succeed in finding $105$ solutions that match the trivial lower bound. 

\paragraph{Literature review}
The \emph{multi-agent path finding} problem (MAPF) has been well studied over the last 20 years. This problem occurs in many industrial applications that involve agents that have to reach destinations without colliding with each other~\cite{ma2017overview}, for instance, in automated warehouses~\cite{li2020lifelong}, autonomous vehicles, and robotics~\cite{bartak2019multi}. Well-studied approaches include search-based solvers (for instance: HCA*~\cite{silver2005cooperative}) that route the agents one by one according to a predefined order. When an agent is routed, it ``reserves'' times and locations. Then, the algorithm routes the next agent until every agent is routed. Such search-based solvers can also route many robots at the same time, thus having for each time-step up to $5^n$ possibilities where $n$ is the number of robots simultaneously routed together (for instance, Enhanced Partial Expansion A*~\cite{goldenberg2014enhanced}). There also exist rule-based solvers that identify scenarios and apply rules (predetermined patterns) to move agents (for instance, the push-and-rotate algorithm~\cite{de2014push}). Some algorithms model the MAPF problem using network flows \cite{icaart17}, integer linear programs~\cite{yu2013planning}, or SAT instances~\cite{surynek2019unifying,DBLP:conf/ecai/SurynekFSB16}.

One of the most popular methods to solve the MAPF problem is the \emph{conflict-based search} (CBS)~\cite{sharon2015conflict}. This method starts by solving a relaxation of the original problem, in which agent collisions are ignored. This relaxation is relatively easy to solve, as it consists of running a shortest-path algorithm for each agent. If the resulting plan contains a time $t$ and coordinate $c$ where two agents $r_1,r_2$ collide, then the algorithm forbids either $r_1$ or $r_2$ from being at the coordinate $c$ at time $t$.  This results in a search tree that is explored until it is depleted (thus finding the optimal solution for the problem). This method has been reported to achieve excellent results and multiple improvements have been made.
Some examples of improvements include a better estimate of the remaining cost~\cite{felner2018adding}, merge-and-restart and conflict prioritization~\cite{boyarski2015icbs}, some suboptimal variants~\cite{barer2014suboptimal}, and some techniques such as a branch-and-cut-and-price algorithm~\cite{lam2019branch}.
For more information about the multi-agent path finding problem, we refer the reader to surveys about the MAPF variants and instances~\cite{stern2019multi} and about the MAPF solvers~\cite{felner2017search}.

At the beginning of the challenge, we tried some of the aforementioned approaches (notably CBS) to solve the challenge instances. To our surprise, CBS did not perform well. Indeed, the challenge instances are much denser than the ones in the literature. The classical MAPF instances are usually sparse in the number of agents (there are from 2 to 120 agents placed in grids with over 100,000 cells in the classical instances~\cite{stern2019multi}, while challenge instances contain hundreds or thousands of agents placed in grids never larger than $100\times100$). This structural difference has a dramatic effect on the performance of CBS in our experiments. We tested the open-source MAPF solver libMultiRobotPlanning \footnote{https://github.com/whoenig/libMultiRobotPlanning} with a memory limit of 16GB and it fails to find solutions for most of the challenge instances (except for some instances with less than $50$ robots, namely \texttt{small\_000} and \texttt{small\_free\_000}).

\medskip

In this paper,  we present several new ideas we used in the competition. We implemented a plethora of different techniques to find initial solutions to different kinds of instances, depending mostly on the size and density of the instance, as well as the presence of obstacles. A key element of our approach was to improve these initial solutions into low-makespan solutions. For this task, we introduced a novel heuristic that we call the \emph{Conflict Optimizer}, which may very well be adapted to other optimization problems.
Our code is available on github \url{https://github.com/gfonsecabr/shadoks-robots}. Some of our results in the competition were obtained by running our solvers on an instance for several weeks.

The remainder of the paper is organized as follows. In Section~\ref{s:initial} we consider the problem of obtaining feasible solutions of moderate makespan. These solutions are optimized later on, with the techniques described in Section~\ref{s:improving}. Details on implementing the algorithms are described in Section~\ref{s:engineering}. Section~\ref{s:results} describes the results we obtained for some challenge instances and presents a comparison with the strategies used by other teams. Concluding remarks are presented in Section~\ref{s:conclusion}.

\section{Initial Solutions} \label{s:initial}

Feasibility is guaranteed for the challenge instances since the number of obstacles is finite and every start and target are located in the unbounded region of space.
In this section, we show how to obtain feasible solutions with a moderate makespan. We divide the heuristics in two categories. In Section~\ref{ss:step}, we compute the solution one step at a time, considering multiple robots simultaneously. In Section~\ref{ss:robot}, we compute the solutions one robot at a time.

The heuristics of the first category are not guaranteed to find a solution, but when they do they often find solutions of lower makespan than those of the second category. The algorithms of the second category are guaranteed to find a solution, but the resulting makespan may potentially be high.

\subsection{Step by Step Computation} \label{ss:step}

The problem of finding a solution for coordinated motion planning in a given number of steps can be modeled as an Integer Linear Problem (ILP) or equivalently as a SAT problem (see \cite{surynek2019unifying,DBLP:conf/ecai/SurynekFSB16} and references therein). While applying such an approach is intractable even for small instances, it can be adapted to find an initial solution. The general idea of the \emph{Greedy} solver is to plan only a small number $k$ of steps for the robots such that the overall distance to the targets decreases as much as possible. Then, we move each robot by one step and repeat this procedure until all robots reach their respective targets.

Our ILP model considers a Boolean variable $x_{r, P}$ for each robot $r$ and for each possible path $P$ of length $k$ starting at the position of the robot.
Constraints of having one and only one path per robot and avoiding obstacles and collisions between robots are easily expressed as linear inequalities. The objective function we maximize is the sum of all the variables with weight
\[\weight(r, P) = \Big( \delta_r(p(0)) - \delta_r(p(k)) \Big) \cdot \Big( (\delta_r(p(0)))^2 + 1 \Big),\]
where $p(0)$ and $p(k)$ are the first and the last positions of the path $P$ and $\delta_r(p)$ is the obstacle-avoiding distance from a point $p$ to the target of the robot $r$. The first factor encourages the solution to push the robots towards their targets, since it is better to get closer to the target. The second factor prioritizes moving robots that are farther from their targets, and we add one so that robots that are already at their target position are encouraged to remain there (otherwise, since their weight is zero, they would be free to move to any position).

In practice, we set $k = 3$ and we only perform the first step of the planned moves, so that the robots can \emph{anticipate} the moves of the other robots. Using the CPLEX~\cite{cplex2009v12} library to solve these problems, we can handle instances with up to roughly 200 robots. Note that this Greedy algorithm is not guaranteed to find a solution. For example, it fails to solve instances with \emph{corridors}.
The reason is that two robots may enter an infinite loop pushing each other back and forth inside the corridor, as the weight function gives higher priority to the robot that is currently farther away from its target, which may alternate between the two robots.

\subsection{Robot by Robot Computation} \label{ss:robot}

\begin{figure}[tb]
  \centering
  \includegraphics[scale=.65]{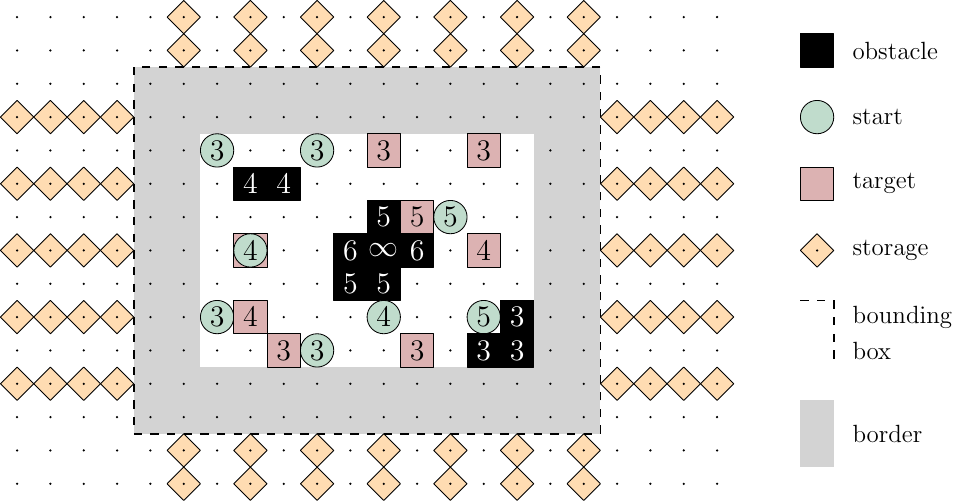}
  \caption{\label{f:storage} Bounding box and storage network, with the depth of start positions, target positions, and obstacles written inside the cells.}
\end{figure}

The algorithms in this section compute the solution one robot at a time using an A* search in two steps. Each robot is assigned an intermediate position called storage and defined later on. First, we use the A* search to compute paths to the storage, one robot at a time. Then, we use A* search again, to compute a path from the initial position to the target position. The order of the robots to compute these paths is an important element that will be detailed later. The path to the intermediate position serves to guarantee that a feasible path exists. Next, we present more details of this approach, and invite the reader to see Figure~\ref{f:storage} for an illustration of the definitions.

We refer to the \emph{bounding box} as an integer axis-aligned rectangular region containing all the start positions, target positions, and obstacles inside its strict interior (not on the boundary).
Given a set of obstacles and a bounding box, the \emph{depth} of a position $p$ is the minimum obstacle-avoiding distance from $p$ to a position outside the bounding box.
One may note that, being in the strict interior of the bounding box, the start and target positions have depth at least two. 
By increasing the size of the bounding box, we can set the minimum depth of the start and target positions, as well as the obstacles to any value $b \geq 2$. We refer to the positions of depth less than $b$ as the \emph{border} of the bounding box.
Indeed the subsequent algorithms require a way for a robot to go around all start and target positions and the border provides such a way.
A larger value of $b$ prevents congestion among robots traveling through the border and may help reduce the makespan. However, setting $b$ to a large value may also increase the makespan by making robots travel too far away from the start position before moving to the target position. A value of $b$ from $2$ to $4$, depending on the algorithm and the instance, seems to be a good compromise in our experience, where instances with more robots tend to benefit from slightly larger values of $b$.

All algorithms in this section are based on a \emph{storage network} $N$. A storage network is a set $N$ of positions outside a predetermined bounding box such that for every position $p$ in $N$, there exists a path that avoids all other positions of $N$ and goes from $p$ to some point in the bounding box. Each robot $r_i$ is assigned to a distinct element of $N$, called the \emph{storage} of $r_i$.

Initially, we set the path of each robot to be stationary at the start position. We sort the robots by \emph{increasing start depth} and for each robot in order, we use A* search to find the shortest path from start to storage, replacing the previous stationary path. The order by which the robots are sorted guarantees that such a path exists.

The A* search happens in 3-dimensional space, where each robot state has integer coordinates of the form $(x,y,t)$ for position coordinates $x,y$ and time $t$. There are $5$ possible \emph{movements}, all of which increase $t$ by one unit. One movement keeps the position $x,y$ unchanged, while the other $4$ movements increment or decrement one of the two coordinates. A movement is feasible if it does not violate any of the problem constraints, considering the currently defined path of the other robots.

After finding paths from start to storage for every robot, we proceed to the next phase of the algorithm. We now sort the robots by \emph{decreasing target depth}. Again, the order of the robots guarantees that a path from storage to target exists. However, we do not compute such a path. Instead, we compute a path from start to target directly, whose existence is guaranteed by the existence of a path from start to storage and another one from storage to target.
The following paragraphs describe the design of four different storage networks.

\begin{figure}[tb]
  \centering
  \includegraphics[scale=.25]{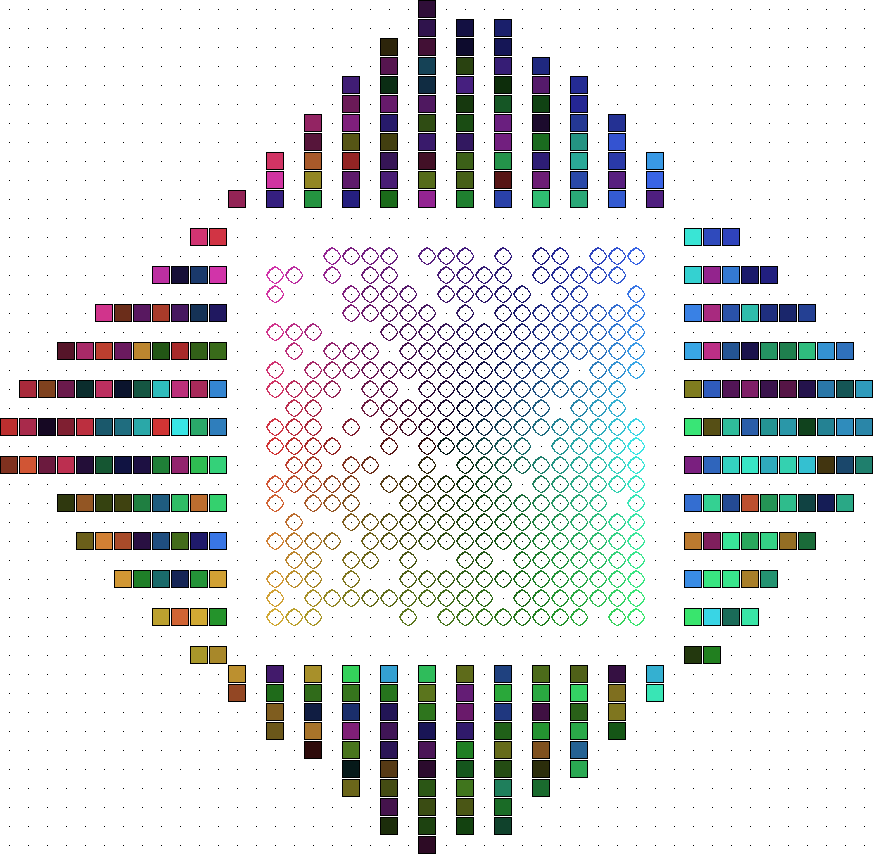}\hspace{.5cm}
  \includegraphics[scale=.25]{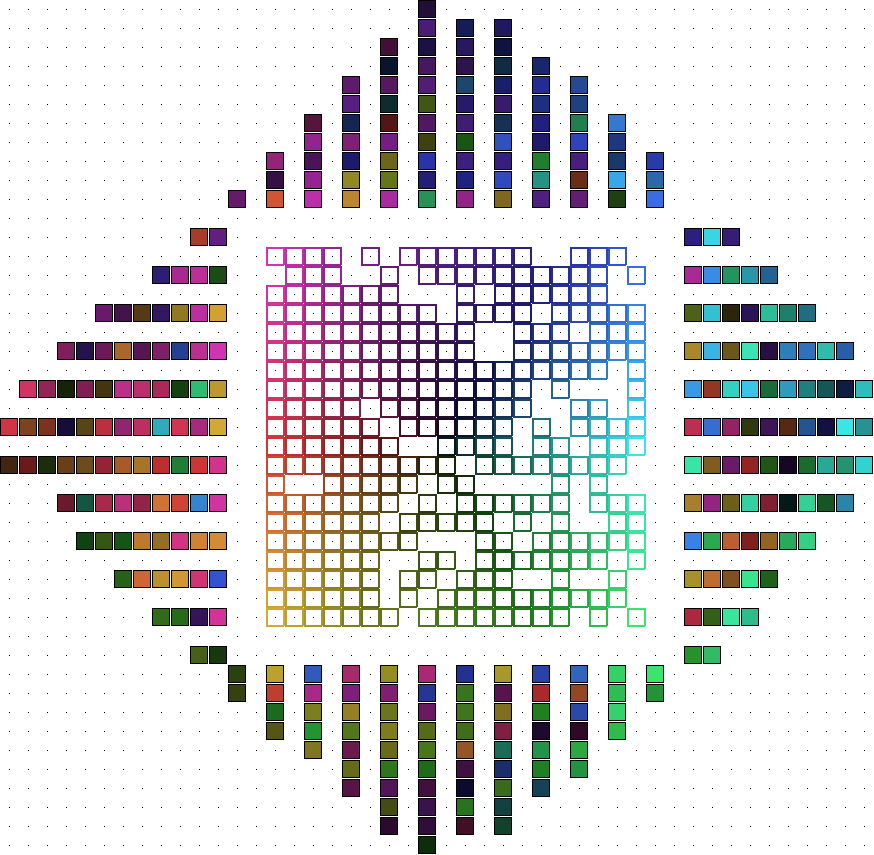}
  \caption{\label{f:cross} Cross storage network for the \texttt{small\_free\_016} instance colored based on start and target locations, respectively.}
\end{figure}

\paragraph{Cross.}
In the \emph{Cross} strategy, we define the storage network $N$ as the set of columns of even $x$ coordinate lying directly above or below the bounding box and the set of rows of even $y$ coordinate lying directly to the left or right of the bounding box, hence the name Cross. Then, we compute a maximal cardinality matching between the robots and $N$. We tried both minimum-weight matching and greedy matchings, minimizing a weight function that considers the distance from start to storage as well as the distance from storage to target. In the greedy matching version, robots are assigned a storage ordered by decreasing start-to-target distance. The result is represented in Figure~\ref{f:cross}. Start positions are represented by a hollow circle, target positions are represented by a hollow square and the storage positions are represented by a slanted solid square. Each robot is assigned a different color following a rainbow pattern based on either the start or the target position to illustrate how the storage assignments are made.

\begin{figure}[t]
  \centering
  \includegraphics[scale=.25]{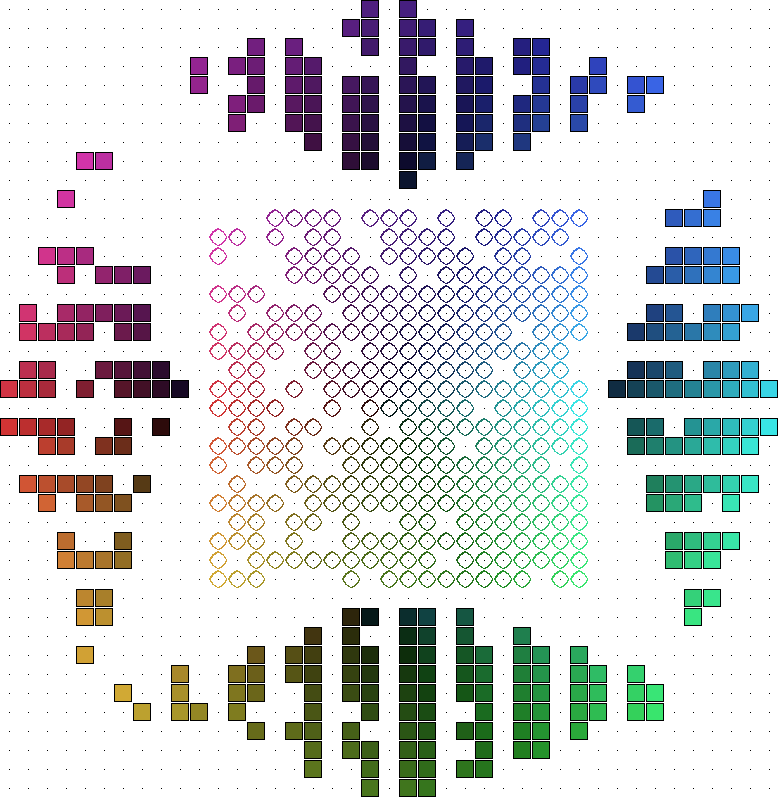}\hspace{.5cm}
  \includegraphics[scale=.25]{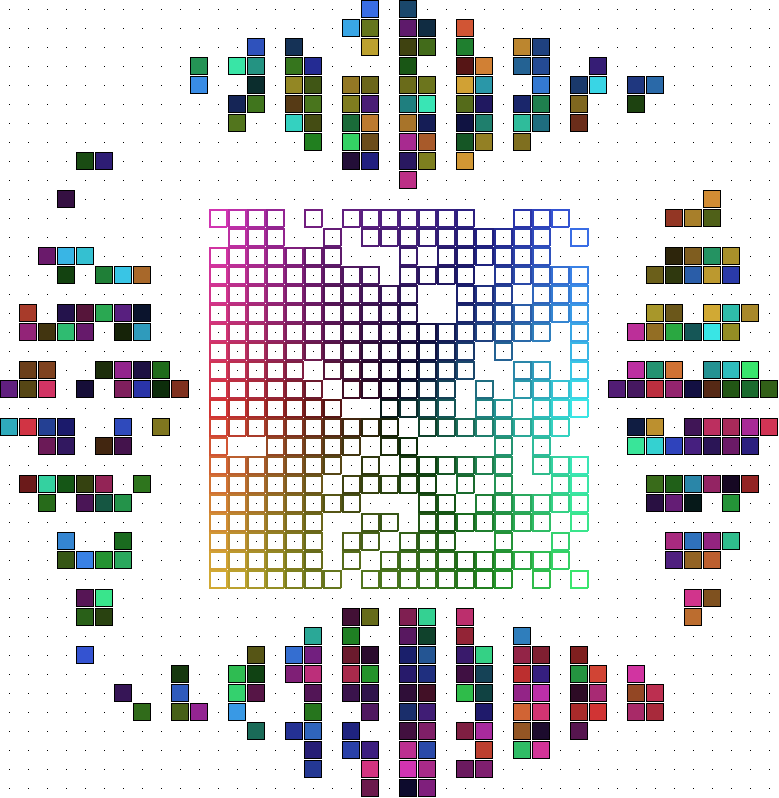}
  \caption{\label{f:cootie} Cootie Catcher storage network for the \texttt{small\_free\_016} instance colored based on start and target locations, respectively.}
\end{figure}

\paragraph{Cootie Catcher.}
The previous strategy works very well for small or sparse instances. However, the different directions of the flow of robots from start to storage make the solutions inefficient for large and dense instances. The \emph{Cootie Catcher} strategy computes the storage using only the start location, in order to better exploit parallel movement of the robots. The storage network shape consisting of four diamonds is presented in Figure~\ref{f:cootie}. When applied to instances without obstacles, the strategy is guaranteed to find a path from start to storage using at most $w/2 + \O(1)$ steps, where $w$ is the largest bounding box side. Surprisingly, this strategy also works well for many instances with obstacles.

\paragraph{Dichotomy.}
The weakness of the previous method is that robots may be assigned storage in a location that is opposite to the direction from start to target. Furthermore, the parallel movement of the robots makes it unlikely that a robot will be able to take any significant shortcuts before it reaches the storage. In order to exploit parallel movements while taking the target location into consideration, we developed the \emph{Dichotomy} strategy. The strategy only works for instances without obstacles.

\begin{figure}[htb]
  \centering
  \includegraphics[scale=.25]{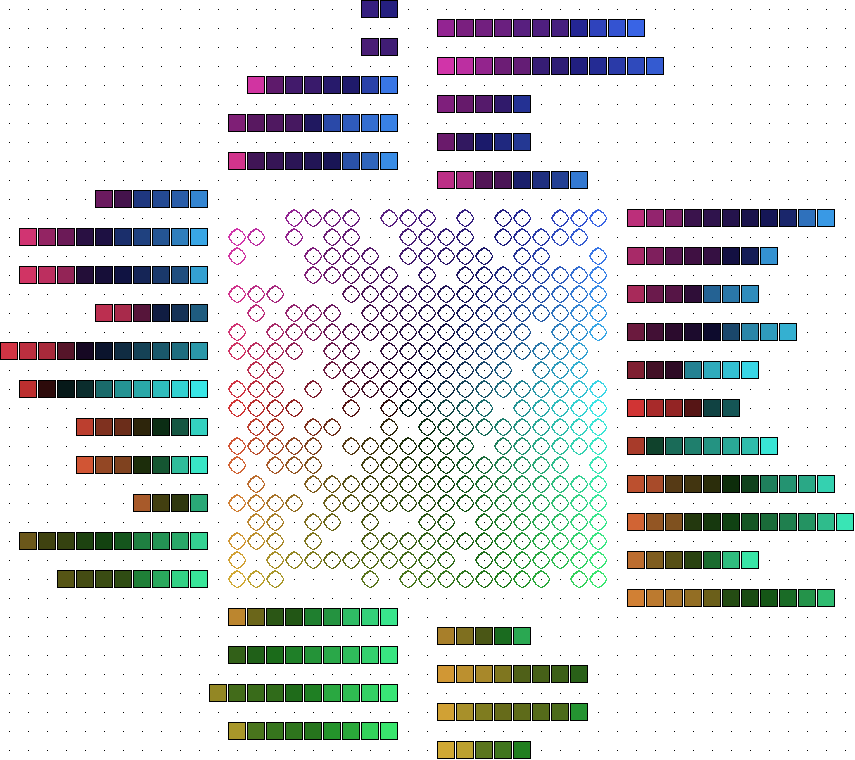}\hspace{.5cm}
  \includegraphics[scale=.25]{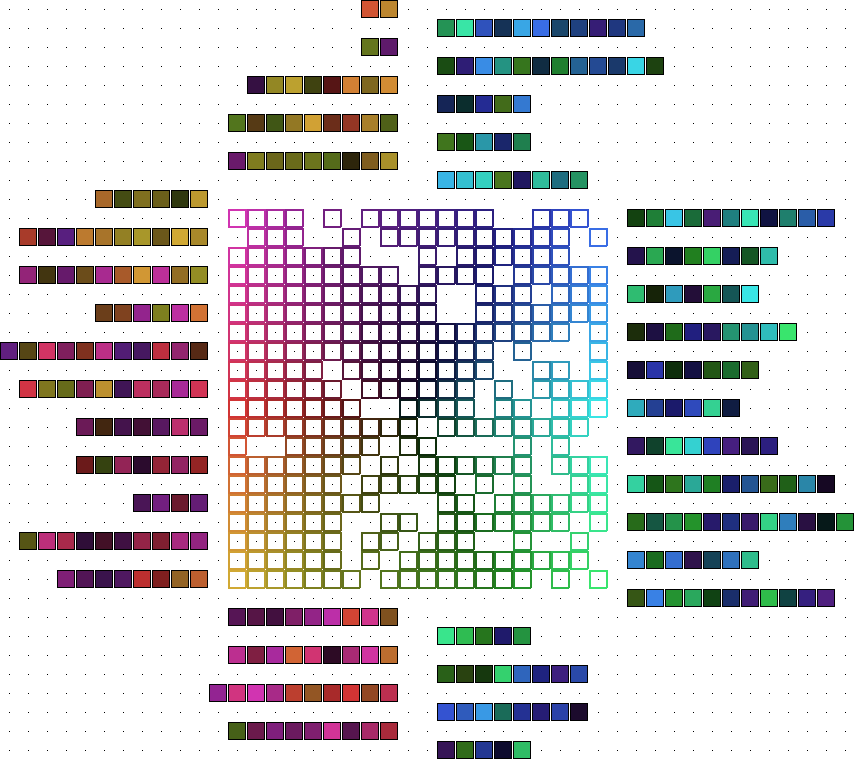}
  \caption{\label{f:dichotomy} Dichotomy storage network for the \texttt{small\_free\_016} instance colored based on start and target locations, respectively.}
\end{figure}

We translate the coordinate system so that the origin is the center of the bounding box. The robots are partitioned into two sets called \emph{left side} and \emph{right side} according to the sign of the $x$-coordinate of the target location. Left-side robots are assigned storage with positive $x$-coordinate while right-side robots are assigned storage with negative $x$-coordinate, as represented in Figure~\ref{f:dichotomy}.

The algorithm performs the following steps, described only for the robots with non-negative $y$-coordinate for simplicity, as the other half is analogous.

\begin{enumerate}
    \item Each robot goes up from start position $(x,y)$ to position $(x,2y)$.
    \item If the robot target is on the right side, the robot moves up one more row. At this point, the even $y$ rows contain left-side robots and the odd $y$ rows contain the right-side robots.
    \item If a right-side (resp. left-side) robot is still inside the bounding box, then it moves to the right (left) as far as needed to leave enough space for the other robots to its left on the same row to move out of the bounding box. Otherwise, a right-side (left-side) robot moves right (left) in order to leave enough space for the robots on the same row to move to a position of positive (negative) $x$-coordinate.
\end{enumerate}

Going from start to storage takes at most $3w/2 + \O(1)$ movements for a $w \times w$ bounding box. Instead of sorting the robots by decreasing target depth as usual, we sort the robots by absolute value of the target $x$-coordinate and then determine the paths from start to target using A* as usual.

\paragraph{Escape.}
This strategy focuses on instances with obstacles, especially on dense instances where the obstacles create bottlenecks to the passage of the robots. 
The goal of the \emph{Escape} strategy is to move all robots outside the bounding box as quickly as possible. To do so, we move the robots by blocks (a \emph{block} is a large polyomino of robots), making efficient use of parallel movements.

\begin{figure}[htb]
  \centering
  \includegraphics[scale=.7]{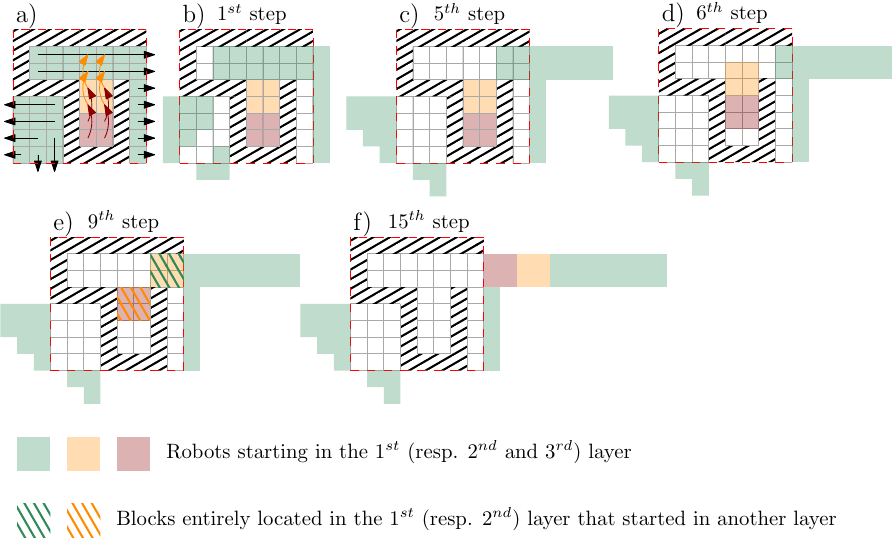}
  \caption{\label{f:escape_strat} Escape strategy. a) In green (resp. orange and red), the first (resp. second and third) layer. The orange arrows show where the second layer will move, the red arrows show the same for the third layer. b-c) The robots in the first layer are moving, but the robots of the second and third layer are still stuck. d) The robots from the second layer are now free to move towards the first layer, this also allows the robots from the third layer to move. e) The robots that used to be located in the third layer, are now located in the second layer and are waiting for their path to be clear of robots. f) Final placement outside the bounding box.}
\end{figure}

The Escape strategy divides the non-obstacle grid cells inside the bounding box into \emph{layers} and each layer is partitioned into blocks.
The first layer is defined so that the grid cells located in it can reach the outside of the bounding box by following a straight line. To make sure that there are no intersections between any two paths taken by robots located in the first layer, the layer is divided in blocks. The robots in each block move at the same time in parallel.
Then, the second layer is defined. It consists of blocks adjacent to the first layer, ideally containing as many cells as possible, that will move in a straight line to the first layer. Again, all robots located in the same block will move towards the first layer in parallel motion. 
Then, in the same way, a third layer is defined, consisting of blocks that will move into the first or second layer. Layers are added until every robot is located in a layer, as represented in Figure~\ref{f:escape_strat}.

\begin{figure}[htb]
  \centering
  \includegraphics[height=6.3cm]{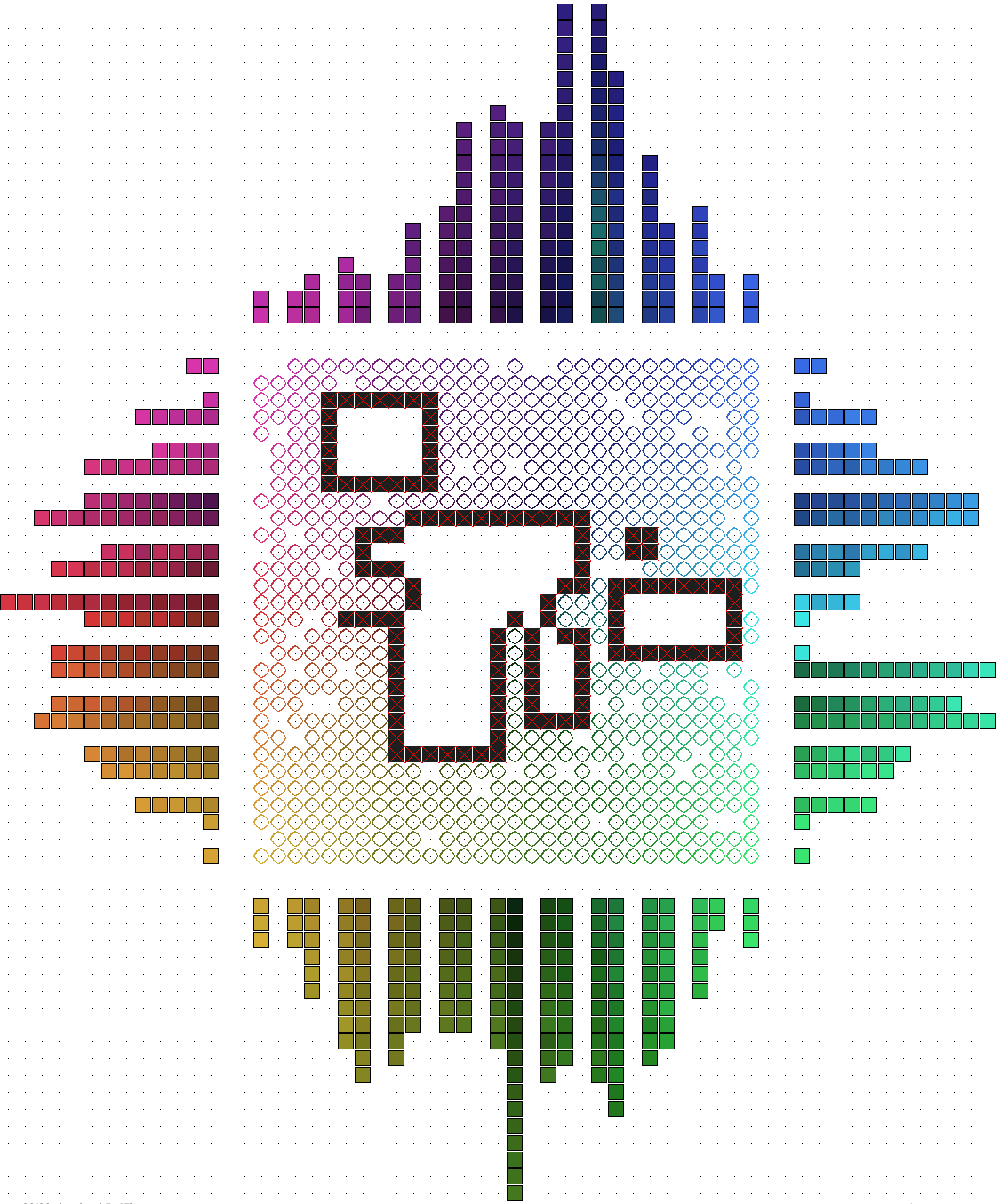} \hspace{.5cm}
  \includegraphics[angle=90,height=6.3cm]{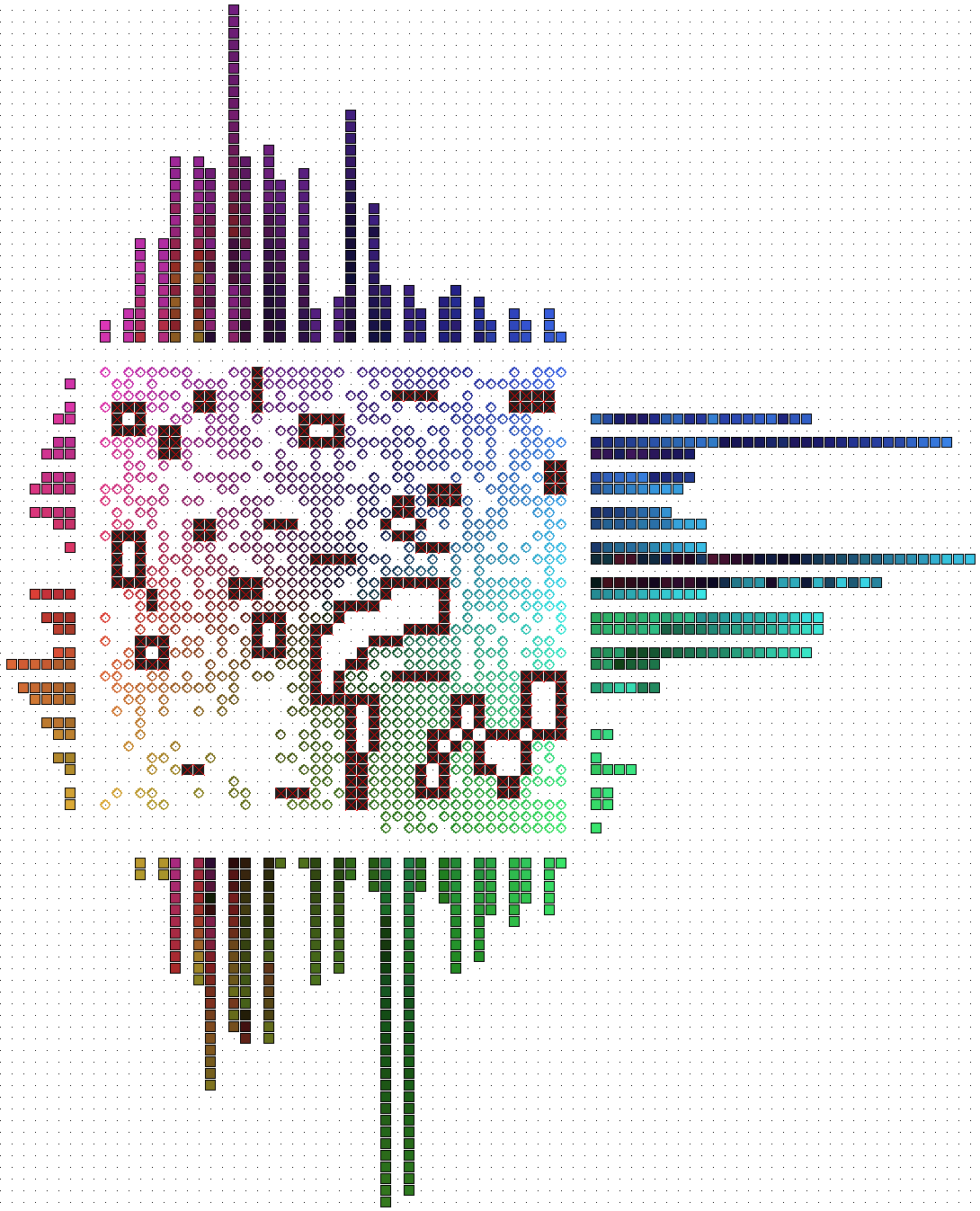}
  \caption{\label{f:escape} Escape storage network for the \texttt{medium\_007} and \texttt{buffalo\_003} instances, respectively.}
\end{figure}

We used a greedy algorithm to define the layers. To compute layer $k$, the algorithm iteratively looks at each block $b_i$ of the previous layers and looks for the largest block not yet assigned to a layer that can move to $b_i$ in a straight line. We partially redefined the layers manually for the most complicated instances and the ones where the algorithm produced unsatisfying results.
As shown in Figure~\ref{f:escape}, outside the bounding box, only two out of every three consecutive rows and columns are used for storage in order to create traveling corridors for the future movements. 

\section{Improving Solutions} \label{s:improving}

In this section, we discuss the two heuristics that we used to reduce the makespan of a given feasible solution. The first heuristic makes local changes to the solution, which remains feasible throughout the process, and possibly reduces the makespan. The second heuristic destroys the feasibility of the solution and either finds another solution of reduced makespan, or no feasible solution at all. Throughout, let $m$ be the makespan of the input solution. 

\paragraph{Feasible Optimizer.}
The idea of the \emph{Feasible Optimizer} is the following. We iteratively remove the path of a robot $r$ from the solution, and then use the A* algorithm to find a new (hopefully different) path for $r$. The A* algorithm may be tuned in several ways to produce different paths, and we do so in such a way that the makespan of the solution never increases and also that a robot is only allowed to move at time $m$ if it already did so in the original path. This way, not only the makespan but also the number of robots moving at time $m$ never increase.  Next, we list some examples on how to modify the A* search.

\begin{itemize}
    \item Find the path from start to target that reaches the target as quickly as possible but break ties using the sum of random weights given to each grid cell the robot passes through.
    \item Reversing the direction of time and then finding a path from target to start that reaches the start as quickly as possible. In the original time direction, that means that the robot will remain at the start for as long as possible.
    \item In the reversed case, force the robot to stay at target for a certain number of steps.
\end{itemize}

\paragraph{Conflict Optimizer.}
The previous optimization strategy may take very long to reduce the makespan because it relies on chance to move a robot away from the path another robot would rather take. Next, we describe a more aggressive approach that leaves the feasible solution space and works far better than we expected. The algorithm uses a modified A* search that allows for a robot to go over another robot's path, which we call a \emph{conflict}. We start by creating a \emph{queue} with all the robots that move at makespan time $m$. While the queue is not empty, we repeat the following procedure for a robot $r$ popped from the front of the queue. 

\begin{enumerate}
    \item Erase $r$'s path.
    \item Find a path for $r$ from start to target that arrives no later than time $m-1$ and minimizes the sum of the weights of the conflicting robots.
    \item Add all conflicting robots to the queue.
\end{enumerate}

Let $q(r)$ be the number of times $r$ has been popped out of the queue. We define the \emph{weight} of a robot $r$ as $1 + (q(r))^2$. This weight function gives some incentive to converge to a feasible solution, preventing robots from repeatedly finding paths that conflict with each other. Notice that $r$'s path is only cleared once it is popped out of the queue. This way, a robot $r$ tries to prevent parts of its previous path from being used by other robots, despite the fact that $r$ will need to find a new path. This little detail makes a big difference, as more likely a big portion of the new path of $r$ will coincide with its previous path.

For sparse or small instances, the Conflict Optimizer can even be used to compute solutions from scratch by choosing an initial makespan and putting all the robots in the queue. This approach fails to find solutions for most instances, though. Hence, we used the Conflict Optimizer only to optimize solutions obtained using the algorithms described in Section~\ref{s:initial}.

\section{Implementation Aspects} \label{s:engineering}

In this section, we describe different techniques used to efficiently implement and apply the previously described heuristics. All heuristics have been executed multiple times, extensively using randomization whenever possible. Furthermore, the problem has several types of symmetry that have been exploited to find more (potentially better) solutions. First, we note that applying rotations (by multiples of 90 degrees) does not change the problem. Second, we note that reversing the start and target positions also does not change the problem. Hence, we applied different rotations as well as reversing the start and the target of the instances.

Multiple executions were used to produce over ten thousand solution files total.  All solution files were saved with a timestamp on the file name. That allowed us to find initial solutions that would optimize better. Even though we only optimized the solutions for makespan, this large volume of solutions allowed us to obtain solutions with a sufficiently low sum of the distances to obtain the third place in that category. Developing tools to efficiently organize and view all of these solutions was an important part of the team strategy. These tools include a viewer, an editor, and a utility to list and copy solution files meeting certain criteria.

The heuristics have been coded in C++, most of which is available publicly on github\footnote{\url{https://github.com/gfonsecabr/shadoks-robots}}. The tools have been coded in python. We executed the code on several Linux machines, both personal computers and high performance computing clusters at the LIS and LIMOS laboratories.

The A* search is in the heart of most of our heuristics. Hence, a lot of work has been done to improve its performance. Sometimes we used deterministic A*, breaking ties by the coordinates of the position, but more often we used a randomized A* algorithm where ties are broken by the sum of the weights of the positions in the path, which are assigned randomly. The A* algorithm needs a distance function as a lower bound and a collision detection, which are described next.

\paragraph{Distance queries} \label{ss:distance} The A* algorithm is guided by a lower bound to the distance to target. In the case without obstacles, the lower bound we used is simply the $L_1$ distance, which can be calculated in $\O(1)$ time. While this lower bound is still valid for instances with a set $O$ of obstacles, it is inefficient because it does not take the obstacles into account. Instead, we used the obstacle-avoiding $L_1$ distance.

Calculating the obstacle-avoiding $L_1$ distance from scratch is a slow process. Since this computation happens many times during the execution of our heuristics, it is essential to be able to compute it quickly. To this purpose, we need to use a data structure to compute the distance $\dist(p)$ from a query point $p$ to a fixed target (in our case, given at preprocessing time). Existing data structures for the problem~\cite{CIW14,ChWa13} seem hard to implement. Instead, we designed a simple data structure that takes $\O(\log w)$ query time for obstacles inside a square of side $w$. The storage requirement may potentially be close to $w^2$, but in our case it is significantly less, generally close to $\O(|O|)$.

\begin{figure}[htb]
  \centering
  \includegraphics[scale=.65]{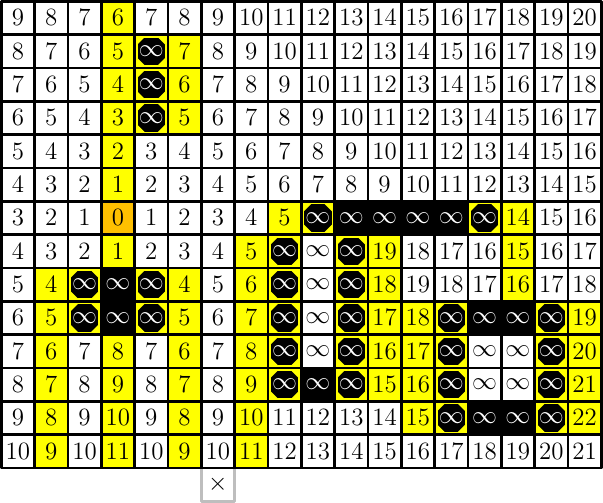}
  \caption{\label{f:dist} Distance with obstacles. Only yellow positions are stored. Black regions represent obstacles and the corners are painted yellow if the value of infinity is stored.}
\end{figure}

Given two consecutive points $(x,y),(x+1,y)$ on the same line we have $\dist(x,y)-\dist(x+1,y) \in \{-1,0,1\}$. Furthermore, if $x$ is to the left (resp., right) of the bounding box, then $\dist(x,y)-\dist(x+1,y) = 1$ (resp. $= -1$). Hence, for each line $y$ in the bounding box we only store the points $(x,y)$ such that $\dist(x,y) \neq (\dist(x-1,y)+\dist(x+1,y))/2$, as shown in Figure~\ref{f:dist}. All the remaining queries for line $y$ can be calculated by interpolating or extrapolating these stored values, which can be located in $\O(\log w)$ time using binary search on a sorted vector.

Queries for a point $(x,y)$ above or below the bounding box are answered by using the closest line of the bounding box and the fact that $\dist(x,y) = \dist(x,y') + |y-y'|$. For example, to query the square marked by a $\times$ in Figure~\ref{f:dist}, we would add $1$ to go one line above, and then interpolate between $9$ and $11$ to obtain $1 + (9+11)/2 = 11$.

\paragraph{Collision detection} \label{ss:collisions}
Fast collision detection between two robots is a key point to the performance of the A* algorithm. We used an internal storage Hopscotch hash table implemented by Thibaut Goetghebuer-Planchon and distributed under the MIT license~\cite{Hopscotch}. Given a position and time, we stored the robot in that position (or the list of robots in that position for the Conflict Optimizer). Hence, collision detection reduces to a small number of hash table lookups.

\paragraph{Avoiding Conflict Optimizer stalls} \label{ss:stall}
The Conflict Optimizer is arguably the most significant contribution of this work. However, it may stall at sub-optimal solutions. To reduce this problem, we may use the following approaches. (i) Reverse start and target as well as the paths in the solution. (ii) Use the Feasible Optimizer to shuffle the solution. (iii) Use randomized paths in the A* search. (iv) Insert the robots that conflict with the same robot in the queue using a random order.

\section{Experimental Results} \label{s:results}

Tables~\ref{tab:free} and~\ref{tab:obs} show the makespan obtained using different heuristics on some selected challenge instances and the makespan lower bound. The Feasible Optimizer column corresponds to the best optimization it obtained starting from different solutions. The Conflict Optimizer column corresponds to the optimization of the solution obtained by the Feasible Optimizer.

\begin{figure}[t]
  \centering
  \includegraphics[scale=.54]{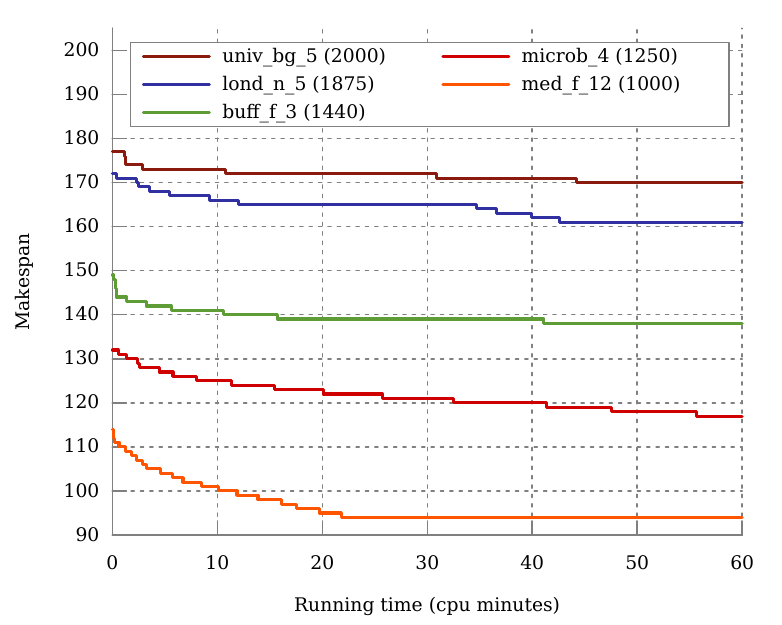}
  \includegraphics[scale=.54]{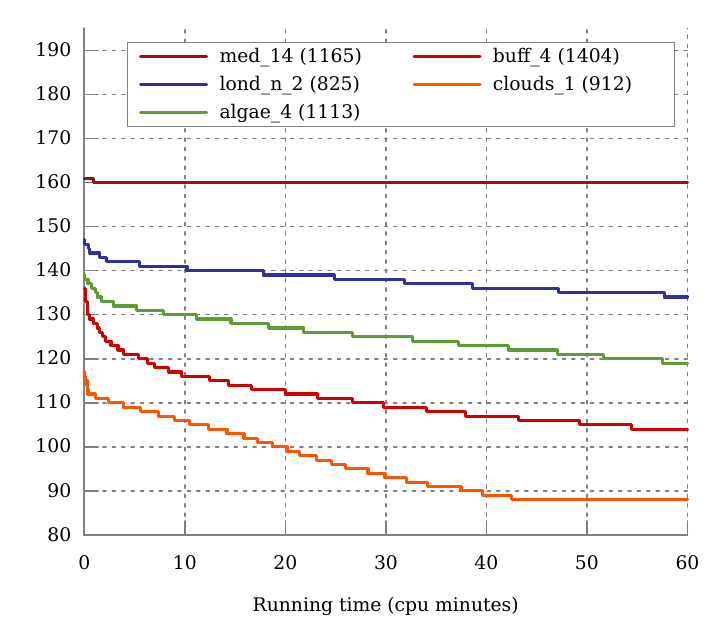}
  \includegraphics[scale=.54]{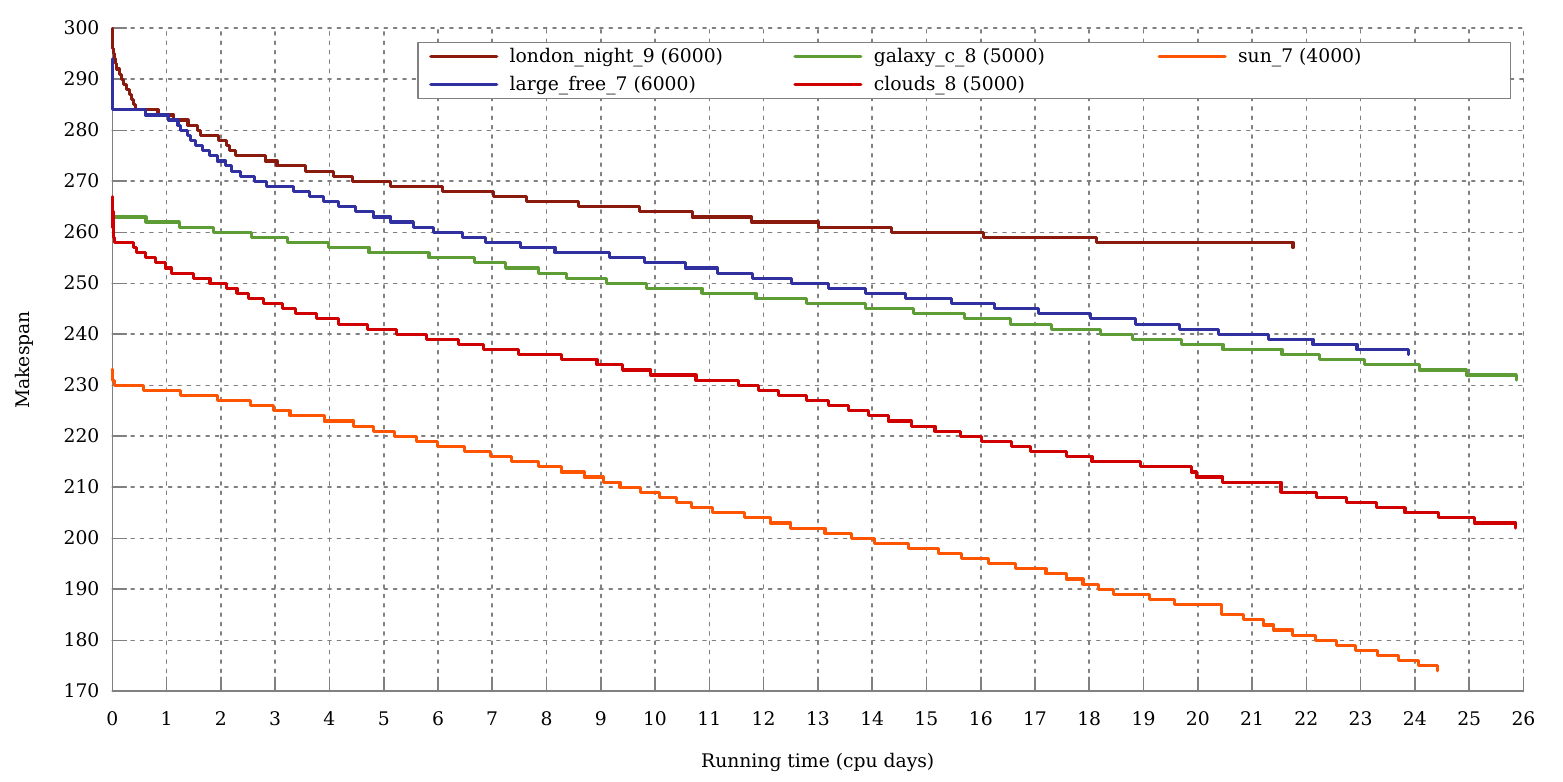}
  \caption{\label{f:conflict} Improved makespan over computation time using the Conflict Optimizer.}
\end{figure}

Figure~\ref{f:conflict} shows the improvement obtained by the Conflict Optimizer over one hour of execution as well as over several weeks. In the case of one hour of execution, we did not attempt to avoid the conflict optimizer stalls. However, the data represented at the bottom of the figure has been produced during the challenge, and is subject to both automatic and manual attempts to avoid stalls. We note that near the end of the challenge, some solutions kept improving very slowly with the Conflict Optimizer: some instances with a few thousand robots such as \texttt{sun\_007}, \texttt{clouds\_008}, and \texttt{large\_free\_007} were consistently giving $1$ unit of makespan improvement for every 10 to 20 hours of computation throughout weeks.

\begin{table}[p]
\centering
\begin{scriptsize}
\begin{tabular}{lrr|rrrr|rr|r}
 & & & \multicolumn{4}{c|}{initial solution} & \multicolumn{2}{c|}{optimizer} & lower  \\
instance                            & $n$  & $w$  & Greedy     & Cross  & Cootie C. & Dichotomy & Feasible & Conflict & bound  \\
\hline
\texttt{small\_free\_002}           & 40   & 10   & \textbf{17}         & 22    & 27     & 22    & 17      & \textbf{15}      & \textbf{15}  \\
\texttt{small\_free\_003}           & 70   & 10   & \textbf{20}         & 31    & 27     & 26    & 20      & \textbf{16}      & 14  \\
\texttt{small\_free\_010}           & 200  & 20   & \textbf{34}         & 46    & 54     & 45    & 33      & \textbf{32}      & \textbf{32}  \\
\texttt{small\_free\_015}           & 280  & 20   & $\cdot$         & \textbf{60}  & 68     & 65    & 51      & \textbf{40}      & 32  \\
\texttt{small\_free\_016}           & 320  & 20   & \textbf{63}         & 68    & 77     & 68    & 60      & \textbf{47}      & 36  \\
\texttt{medium\_free\_007}          & 630  & 30   & 148        & \textbf{89}    & 103    & 95    & 81      & \textbf{60}      & 52  \\
\texttt{medium\_free\_009}          & 800  & 40   & \textbf{93}        & 97    & 124    & 109   & 81      & \textbf{71}      & \textbf{71}  \\
\texttt{medium\_free\_012}          & 1000 & 50   & $\cdot$    & \textbf{114}   & 125    & 127   & 96      & \textbf{94}      & \textbf{94}  \\
\texttt{microbes\_004}              & 1250 & 50   & $\cdot$    & \textbf{132}   & 159    & 135   & 125     & \textbf{91}      & \textbf{91}  \\
\texttt{buffalo\_free\_003}         & 1440 & 60   & $\cdot$    &\textbf{149}  & 165   & 158     & 125      &  \textbf{87}  & 78\\
\texttt{london\_night\_005}         & 1875 & 50   & $\cdot$    & 179   & 190    & \textbf{173}   & 157     & \textbf{124}     & 92  \\
\texttt{universe\_bg\_005}          & 2000 & 50   & $\cdot$    & 194   & 198    & \textbf{177}   & 173     & \textbf{141}     & 82  \\
\texttt{galaxy\_c2\_008}      & 3000 & 100  & $\cdot$    & \textbf{198}   & 258    & 234   & 168     & \textbf{163}     & \textbf{163} \\
\texttt{large\_free\_004}           & 3938 & 75   & $\cdot$    & 274   & 276    & \textbf{256}   & 240     & \textbf{204}     & 127 \\
\texttt{large\_free\_005}           & 5000 & 100  & $\cdot$    & \textbf{260}   & 316    & 293   & 252     & \textbf{184}     & \textbf{184} \\
\texttt{large\_free\_007}           & 6000 & 100  & $\cdot$    & \textbf{297}   & 343    & 325   & 295     & \textbf{236}     & 189 \\
\texttt{sun\_009}                   & 7500 & 100  & $\cdot$    & 424   & 395    & \textbf{361}   & 354     & \textbf{345}     & 187 \\
\texttt{large\_free\_009}           & 9000 & 100  & $\cdot$    & 514   & 440    & \textbf{391}   & 378     & \textbf{374}     & 182 \\
\end{tabular}
\end{scriptsize}
\caption{Makespan of different heuristics for selected instances without obstacles.}
\label{tab:free}
\end{table}

\begin{table}[p]
\centering
\begin{scriptsize}
\begin{tabular}{lrr|rrrr|rr|r}
 & & & \multicolumn{4}{c|}{initial solution} & \multicolumn{2}{c|}{optimizer} & lower  \\
instance                   & $n$  & $w$  & Greedy   & Cross & Cootie C. & Escape & Feasible & Conflict          & bound  \\
\hline
\texttt{small\_005}                 & 63   & 10   & \textbf{27}     & 28    & 32     &37& 25      &  \textbf{20}     & 18  \\
\texttt{sun\_000}                 & 143  & 20   & \textbf{32}     & 39    & 46     & 61& 29      &  \textbf{27}     & \textbf{27}  \\
\texttt{small\_011}                 & 183  & 20   & \textbf{56}     & 60    & 70     & 67     & 48      &  \textbf{40}     & 37  \\
\texttt{small\_016}                 & 276  & 20   &$\cdot$& \textbf{67}    & 72     & 79     & 57      &  \textbf{43}     & 36  \\
\texttt{medium\_005}                & 407  & 30   &$\cdot$& 119   & 110    & \textbf{106}     & 94      &  \textbf{74}     & 58  \\
\texttt{london\_night\_002}       & 825  & 50   &$\cdot$& \textbf{149}   & 162    & 165    & 142     & \textbf{94}      & 84  \\
\texttt{microbes\_002}            & 958  & 50   &$\cdot$& \textbf{111}   & 135    & 173 & 97      & \textbf{89}      & \textbf{89}  \\
\texttt{clouds\_001}              & 912  & 50   &$\cdot$& \textbf{117}   & 138    & 159 & 94      & \textbf{83}      & \textbf{83}  \\
\texttt{medium\_014}                & 1165 & 40   &$\cdot$& 180   & \textbf{161}    & 180    & 161     & \textbf{151}     & 73  \\
\texttt{algae\_004}               & 1113 & 50   &$\cdot$& \textbf{139}   & 160    & 191 & 121     & \textbf{84}      & 79  \\
\texttt{buffalo\_004}               & 1404 & 60   &$\cdot$& \textbf{136}   & 164    & 195    & 120     & \textbf{104}     & \textbf{104}  \\
\texttt{large\_003}                 & 1906 & 100  &$\cdot$& \textbf{172}   & 224    & 250    &\textbf{154}& \textbf{154}      & \textbf{154}  \\
\texttt{large\_004}                 & 2034 & 100  &$\cdot$& 431   & 391    & \textbf{381} &$\cdot$ & \textbf{381}     & 185 \\
\texttt{large\_005}                 & 3223 & 75   &$\cdot$& 398   & \textbf{310}    & 317 &$\cdot$ & \textbf{299}     & 141  \\
\texttt{universe\_bg\_007}        & 3820 & 100  &$\cdot$& \textbf{224}   & 289    & 323    & 202     & \textbf{184}     & \textbf{184}  \\
\texttt{large\_007}                 & 4706 & 100  &$\cdot$& 753   & 497    & \textbf{491}    & 497     &  \textbf{471}    & 215  \\
\texttt{microbes\_008}            & 5643  & 100 &$\cdot$& \textbf{329}   & 359    & 425    & 322     & \textbf{279}     & 188  \\

\texttt{algae\_009}               & 7311  & 100 &$\cdot$& 500   & \textbf{439}    & 441    & $\cdot$& \textbf{421}     & 176  \\

\texttt{large\_009}                 & 8595  & 100 &$\cdot$& 398   & \textbf{387}    & 566    & $\cdot$& \textbf{352}     & 176  \\

\end{tabular}
\end{scriptsize}
\caption{Makespan of different heuristics for selected instances with obstacles. Note that the Dichotomy algorithm does not work with obstacles and has been replaced by the Escape algorithm.}
\label{tab:obs}
\end{table}

\paragraph{Comparison with other teams}
The two other teams \texttt{UNIST}~\cite{UNIST} and \texttt{gitastrophe}~\cite{gitastrophe} on the podium of the CG:SHOP 2021 challenge used a two-phase strategy similar to ours: first compute an initial solution and then optimize it.

The other two teams also used initial solutions computed through a storage network. The existence of a solution is also guaranteed by using the depth of the start and target positions. We noticed that \texttt{gitastrophe} used the same trick we did: first compute a partial solution going from start to storage and then compute new paths going from start to target since the existence of such paths is guaranteed. \texttt{gitastrophe} also used a minimum weight matching to assign each robot to a storage position. The main difference between the teams during this phase is in the choice of the storage network 
(points within pairwise $L_\infty$ distance at least $2$ for \texttt{gitastrophe} and points having even coordinates for \texttt{UNIST}). Among the different storage networks that we used (Cross, Cootie Catcher, Dichotomy, Escape), we noticed that none of the four is always the best one. 

For the optimization of an initial solution, there are again some similarities, but the differences are more significant. The standard strategy is to randomly remove the paths of a sample of robots before recomputing them with the hope of an improvement. \texttt{gitastrophe} experimented with different ways to choose the samples: according to their makespan, relative distance, or conflicts with a given robot. \texttt{UNIST} used a sample of only one robot as we did in our \emph{Feasible Optimizer} but \texttt{UNIST} also added an original simulated annealing optimization step. Neither \texttt{UNIST} nor \texttt{gitastrophe} have used an optimization algorithm where the sample of robots to recompute evolves dynamically as in our \emph{Conflict Optimizer}. The reason could be that this strategy destroys the solution feasibility without any guarantee of recovery. However, the Conflict Optimizer is probably the main ingredient that gave us a dramatic advantage over the other teams.

\section{Conclusion and Perspectives} \label{s:conclusion}

We developed several algorithms to solve the coordinated motion planning instances of the CG:SHOP2021 challenge, obtaining low-makespan solutions. Considering how dense the instances are, we were surprised that our algorithms found optimal solutions (matching the trivial lower bound obtained by routing each robot independently) for 105 out of 203 challenge instances. Furthermore, our team obtained the best solution found for 202 out of 203 challenge instances.

We attribute this success to the large variety of algorithms that we used. The 3-step \emph{Greedy} algorithm cannot solve most instances, but when it does, it provides solutions that are better than the robot by robot solutions. In the robot by robot solutions, we used several different kinds of storage networks, adapting to the peculiarities of each instance. Besides the four storage networks described herein, a few others have been tested, including the possibility of storage inside the bounding box, as well as restricting the movement direction in chosen cells in order to create one-way lanes. These other networks did not yield any noticeable benefit, though. Optimizing the solutions is a key ingredient. The \emph{Feasible Optimizer} helps, but about half of the optimal solutions that we found were produced by the \emph{Conflict Optimizer}. We were surprised how well this algorithm works, while not surprised that the higher the robots density is, the lower is its efficiency. 

We should keep in mind that the algorithms to find initial solutions have been designed for the challenge instances. A natural question is to determine if the strategy that we used remains suitable for other types of instances, such as the ones arising in the following three scenarios:
\begin{enumerate}
    \item The start and the target positions are located in different areas. For example, consider an instance with the start positions in the square $[-size, -1] \times [-size, size]$ and the target positions in $[1, size] \times [-size, size]$, where $size$ is a parameter determining the size of the different storage spaces. We separate the start and the target positions with a vertical line of obstacles at coordinates $(0,y)$ for $|y|>door$, where $door$ is a parameter determining the dimension of the corridor connecting them.
    This scenario corresponds to the practical problem of moving the content of a warehouse to another location close by.
    \item A set of obstacles form a bounding box around the start and the target positions. An example of such instance is an instance with obstacles around the square $[0, size] \times [0, size]$ and all the start and the target positions located in this square. This scenario differs from the challenge instances since there is no way to make the robots escape the bounding box. A well-known special case of this family of problems is the 15-puzzle, where 15 robots represented by squares have to be reconfigured in a $4\times 4$ grid. These puzzles have been introduced by Noyes Chapman in 1874 and became a craze in the 1880s in the US~\cite{wiki:15_puzzle}. Mathematical results showing that there does not always exist a solution date back to 1879~\cite{johnson1879}. The computation of the shortest solution for this class of puzzles is NP-hard~\cite{ratner1986}.
    More generally, in addition to its playfulness, this scenario corresponds to the practical problem of reorganizing a warehouse without much free space.
    \item At last, we can imagine a set of obstacles defining a bounding box with a small number of free squares (doors) connecting the interior and the exterior of this area. Start and target positions may be both inside and outside the bounding box. Some of the most difficult instances of the CG:SHOP 2021 challenge were of this kind, with all the start and target positions in the bounding box and with only a few doors to go outside. This scenario corresponds to the practical problem of loading and unloading a warehouse.
\end{enumerate}

A key tool that we used for computing the initial solution is a storage network. Storage networks could potentially be used in scenarios 1 and 3, but may be very inefficient for scenario 3 due to a bottleneck at the doors. In scenario 2, it may be possible to adapt the storage network idea, as long as the density is low enough.
Tackling these theoretical problems is a possible direction for future works. Another direction is to fully go to practice by addressing concrete applications of MAPF in warehouses managed by Kiva or Quictrons robots systems. The combinatorial problems of order picking in Robotic Mobile Fulfillment Systems are different from the classical MAPF. The start and target positions of the robots are not necessarily constrained but the robots have to pass through a picking area, for filling the order streaming without wasting time~\cite{valle2021,Liu2019,boysen2017,rimele2021,rimele2021-2}. Efficient heuristics for MAPF are surely a part of the solution but playing with the other variables than the paths opens many other perspectives. 

The Conflict Optimizer, however, is not restricted to the coordinated motion planning problem. For example, we can use the same strategy to optimize the vertex coloring of a graph. Given a graph $G=(V,E)$ and a $k$-coloring $c:V \rightarrow \{1,\ldots,k\}$, we initialize the queue with the vertices of color $k$. Then, until the queue becomes empty, we pop a vertex $v$ from the queue and color it with a color at most $k-1$ while minimizing the sum of weight of the conflicting vertices and adding the conflicting vertices to the queue. If the queue is empty, then we succeeded in coloring the graph using $k-1$ colors. This strategy is one of the main strategies used by the \texttt{Shadoks}, \texttt{Gitastrophe}, and \texttt{LASAOFOOFUBESTINNRRALLDECA} teams to respectively win first to third places in the CG:SHOP 2022 challenge, which consists of vertex-coloring the intersection graphs of line segments in the plane.

\section{Acknowledgments}

We would like to thank Hélène Toussaint, Raphaël Amato, Boris Lonjon, and William Guyot-Lénat from LIMOS, as well as the Qarma and TALEP teams and Manuel Bertrand from LIS, who continue to make the computational resources of the LIMOS and LIS clusters available to our research. We would also like to thank the challenge organizers and other competitors for their time, feedback, and making this whole event possible.

The work of Loïc Crombez has been sponsored by the French government research program ``Investissements d'Avenir'' through the IDEX-ISITE initiative 16-IDEX-0001 (CAP 20-25).
The work of Guilherme D. da Fonseca is supported by the French ANR PRC grant ADDS (ANR-19-CE48-0005).
The work of Yan Gerard is supported by the French ANR PRC grants ADDS (ANR-19-CE48-0005), ACTIVmap (ANR-19-CE19-0005) and by the French government IDEX-ISITE initiative 16-IDEX-0001 (CAP 20-25).
The work of Aldo Gonzalez-Lorenzo is supported by the French ANR PRC grant COHERENCE4D (ANR-20-CE10-0002).
The work of Pascal Lafourcade is supported by the French ANR PRC grant MobiS5 (ANR-18-CE39-0019), DECRYPT (ANR-18-CE39-0007), SEVERITAS (ANR-20-CE39-0005) and by the French government IDEX-ISITE initiative 16-IDEX-0001 (CAP 20-25).
The work of Luc Libralesso is supported by the French ANR PRC grant DECRYPT (ANR-18-CE39-0007).


\bibliographystyle{abbrv}
\bibliography{robot}

\end{document}